\documentclass[twocolumn,showpacs,preprintnumbers,amsmath,amssymb]{revtex4}

\usepackage{graphicx}
\usepackage{dcolumn}
\usepackage{bm}

\newcommand{\be}{\begin{equation}}

\newcommand{\ee}{\end{equation}}
\newcommand{\ba}{\begin{eqnarray}}
\newcommand{\ea}{\end{eqnarray}}
\newcommand{\bd}{\begin{displaymath}}
\newcommand{\ed}{\end{displaymath}}
\newcommand{\nn}{\nonumber}

\newcommand{\w}{\varpi}

\newcommand{\m}{\dot{M}_+}

\newcommand{\mt}{\widetilde{\dot{M}}_+}

\begin{document}

\title{Two Dimensional Poynting Flux Dominated Flow onto a  Schwarzschild Black Hole}

\author{Hyun Kyu Lee}
\email{hklee@hepth.hanyang.ac.kr}
 \altaffiliation[Also at ]{APCTP, Pohang 790-784, Korea}
\author{Jaehong Park}%
\email{jaehong@ihanyang.ac.kr}
\affiliation{Department of Physics
Hanyang University, Seoul 133-791, Korea}

\date{\today}

\begin{abstract}

We discuss the dynamics of the accretion flow onto a black hole
driven by Poynting flux in a simplified model of a two-dimensional
accretion disk.  In this simplified model, the condition of  the
stationary accretion flow  is found to impose a nontrivial
constraint on the magnetic field configuration. The effect of the
magnetic field on the accretion flow is discussed in detail using
the paraboloidal and hyperboloidal type configuration for the
poloidal structure suggested by Blandford in 1976. It is
demonstrated explicitly that the angular velocity of the disk,
$\Omega_D$,  deviates from the Keplerian angular velocity. The
angular velocity of the rigidly-rotating magnetic surface,
$\Omega_F$,  does not have to be the same as the angular velocity
of the disk for the paraboloidal type configuration.  But for the
hyperboloidal type configuration,  it is found that we can set
$\Omega_F = \Omega_D$, which corresponds to an accretion disk of
perfect conductor.  We discuss the numerical solutions of the
stream equation for  stationary accretion flow in the
Schwarzschild background using a paraboloidal type configuration.
The dynamics of the accretion disk is found to depend strongly on
the ratio of the accretion rate to the magnetic field strength.

\end{abstract}

\pacs{97.10.Gz, 97.60.Lf}

\maketitle

\section{Introduction}

To describe  the powerful  and highly collimated astrophysical
jets observed in AGN and quasars, Poynting flux model has been
suggested long time ago \cite{lovelace,blandford} and many
interesting works  have been developed.  One of the
characteristics of  Poynting flux  is that it carries very little
baryonic component compared to the hydrodynamic flow. Recently
this  property of Poynting flux is found to be consistent with the
required property for powering GRB\cite{piran}.  And the Poynting
flux in a system of black hole-accretion disk has also been
studied in connection with the gamma ray bursts\cite{lwb,lbw,li1}
and it is found that   the evolution of the system is  largely
depend on the Poynting outflow from the disk\cite{lk,li}.

The configuration of the ordered magnetic field around the
accretion disk, which is responsible for the Poynting flux, has
been discussed both in analytical and numerical studies. For the
strong enough electromagnetic field around the compact object, the
force-free magnetosphere can be established.  In the
non-relativistic formulation, Blandford\cite{blandford} suggested
an axisymmetric and stationary solution  for the Poynting outflow,
assuming a force-free magnetosphere surrounding an accretion disk.
The poloidal field configuration for a black hole in a force-free
magnetosphere has been discussed recently by Ghosh\cite{ghosh} in
the relativistic formulation using Grad-Shfranov equation.
 The developments  of the
ordered magnetic field in the disk and the Poynting outflow from
the disk have been studied numerically by many
authors\cite{ga,lop,fendt}. Recently  Ustyugova et al.\cite{love}
performed an axisymmetric magneto-hydrodynamical simulation to
show that the quasi-stationary and approximately force-free
Poynting jet from the inner part of the accretion disk is
possible.

In studying the accretion flow, it is also important to know the
dynamics of the accretion disk influenced by the magnetic field.
The energy and angular momentum carried out by the Poynting flux
are determined essentially by the torque and the angular velocity
of the magnetic surface $\Omega_F$, which depends strongly on the
disk dynamics subjected to the magnetic field.  We adopt a
simplified two-dimensional accretion disk model\cite{hklee}
embedded in a  stationary, axisymmetric and force-free
magnetosphere. To see the magnetic effect transparently it is also
assumed that there is no viscous stress tensor in the disk and
there is no radiative transfer from the disk in this model.

The state of an accretion flow is governed by a set of equations
obtained from the conservation of stress-energy tensor\cite{hklee}
and the force-free magnetosphere with the current configuration is
determined by the Grad-Shafranov equation.  The two sets of
equations are not completely independent to each other because
they are coupled at the disk on the equatorial plane.  Hence the
variables in the Grad-Shfranov equations, the stream function
$\Psi$,  the current potential $I$ and $\Omega_F$ are constrained
by the  accretion equations. In other words, the accretion
equations  provide  a kind of boundary values of $\Psi$, $I$ and
$\Omega_F$  which cannot be determined by Grad-Shafranov equation
alone.  It also impose a nontrivial constraint on the
configuration type of the magnetic field. For example, it is found
that the hyperboloidal configuration suggested by Blandford in
1976\cite{blandford} is consistent with an accretion disk of
perfect conductor, that is $\Omega_F = \Omega_D$, for  the
stationary accretion flow, whereas for the paraboloidal type
configuration $\Omega_F$  does not have to be the same as $
\Omega_D$.

In section 2, we describe  the simplified two-dimensional
accretion disk model dominated by Poynting flux in a force-free
magnetosphere. The system is governed by four basic equations:
three accretion equations and the Grad-Shafranov equation in the
Schwarzschild background. In section 3, numerical analysis of the
accretion flow and the field configuration of paraboloidal type
are discussed in detail for two cases of lower and higher
accretion rates. The results are summarized and discussed in
section 4.

\section{Two dimensional accretion disk in force-free magnetosphere}

In this work, a  black hole-accretion disk system  is supposed to
be surrounded by the ordered electromagnetic field, which is
determined by the electric charge and current distributions  as a
solution of the inhomogeneous Maxwell equation: \ba
F^{\mu\nu}_{\;;\nu} = 4\pi J^{\mu} \ea where $J^{\mu}$ are the
$\mu$-th component of the bulk electromagnetic current density.
The boundary values for the field configuration are responsible to
the surface current densities on the horizon and/or on the
accretion disk. It is a kind of boundary value
problem\cite{damour}, which can be described  by introducing
appropriate surface current densities, $j^{\mu}$, defined  on the
surface such that the total conserved current density,  is given
by
 \begin{equation} {\mathcal J}^\mu=J^\mu + j^\mu
        \label{damour1}. \end{equation}
Using the current conservation,
\begin{equation} {\mathcal J}^\mu_{\;\;;\mu}=0 \label{current3}, \end{equation}
one can identify the surface current density when provided with
the geometry of the surface.

The horizon of a black hole is a mathematically well defined
surface, on which  the appropriate surface current density has
been studied in depth\cite{TPM}.  On the other hand, the surface
of the physical accretion disk is not expected to have any sharp
boundary and various  shapes have been suggested depending on the
properties of accretion flow. In this work, however,  we assume a
simplified accretion disk, which is vertically squeezed down to
the equatorial plane, a two-dimensional accretion disk. Then we
can associate the surface current density in a simple way,
 \be j^\mu={1\over
        4\pi}(F^{\theta\mu}_{+}-F^{\theta\mu}_{-})
        \delta(\theta-\pi/2) \label{surface-current}, \end{equation}
which are nothing but Gauss' law and Ampere's law as given by \ba
\sigma_e &=& -{1\over
        4\pi}(E^{\hat{\theta}}_{+}-E^{\hat{\theta}}_{-})
        \label{bc1},\\
        K^{\hat{r}} &=&-{1\over
        4\pi}(B^{\hat{\phi}}_{+}-B^{\hat{\phi}}_{-}) ,\quad
        K^{\hat{\phi}}={1\over 4\pi}(B^{\hat{r}}_{+}-B^{\hat{r}}_{-})
        \label{bc2},\ea
where $\sigma_e$ and $K^i$ are surface charge and surface current
density(spatial) respectively and   $+/-$ denote the upper/lower
hemisphere.

\subsection{Accretion equations}

The dynamics of an accretion flow is determined  by the
conservation equation of the stress-energy tensor:
 \begin{equation} T^{\mu\nu}_{\quad;\mu}=0.\end{equation}
The stress-energy tensor is decomposed into two parts, the matter
($T^{\mu\nu}_m$) and the electromagnetic ($T^{\mu\nu}_{EM}$)
parts: \ba T^{\mu\nu}=T^{\mu\nu}_m +T^{\mu\nu}_{EM} .\ea The
electromagnetic part is given by \begin{equation} T^{\mu\nu}_{EM}
= {1\over 4\pi} \left( F^\mu_{\;\;\rho} F^{\nu\rho}-{1\over 4}
g^{\mu\nu}F_{\rho\sigma}F^{\rho\sigma} \right) .\end{equation}  In
general the stress-energy tensor for the accretion disk is
determined by mass density, internal energy, pressure, viscosity,
radiative transfer and etc.\cite{acgl}. Since we are interested in
the accretion dominated by the Poynting flux, the  two-dimensional
accretion disk is assumed to be non-viscous, cool, and
non-radiative such that
 the matter
part is given by \begin{equation} T^{\mu\nu}_m =\rho_m u^\mu u^\nu
\label{matter},\end{equation} where $\rho_m$ is the rest-mass
density and $u^\mu$ is the four velocity of the accreting matter.

From the  conservation of stress-energy tensor, one can
obtain\cite{hklee}
    \ba (\partial_r u_0)\dot{M}_+  +2\pi r
K^{\hat{r}} (-\alpha
        E^{\hat{r}}+\beta \w B^{\hat{\theta}})=0, \label{ace} \\
(\partial_r u_\phi) \m +2\pi r K^{\hat{r}}\w
        B^{\hat{\theta}}=0, \label{acl}\\
 {\m\over 2\pi r^2} g^{rr} {u^0\over u^r} (\partial_r u_\phi) \left[
        \Omega_D+{\partial_r u_0 \over
        \partial_r u_\phi} \right] \nonumber \\
 -{\sqrt{\Delta}\over
        r^2}\left(\sigma_eE^{\hat{r}}-K^{\hat{\phi}}B^{\hat{\theta}}\right)
        =0 \label{acr}.\ea
where the angular velocity of the disk is given by $ \Omega_D=
u^\phi/u^0$ and the mass accretion rate is given by \ba \dot{M}_+
= -2\pi r \sigma_m u^r \label{mrate}.\ea Eq.(\ref{ace}) and
(\ref{acl}) correspond to the energy and angular momentum
conservation respectively in the stationary and axisymmetric
setting in this work. One can see that the accretion is determined
by the Poynting flux.  The last equation is obtained from the
radial component which essentially determines the orbital motion
of the disk. Since $u^{\theta}=0$, the $\theta$-component of the
conservation equation does not contribute to the accretion flow.
In the absence of external fields, the angular velocity is
determined as \ba \Omega_D =-{\partial_r u_0 \over
\partial_r u_\phi},  \ea which is one of the characteristics of
the Keplerian orbit.  Hence deviations of $\Omega_D$ from the
Keplerian one  is naturally  expected in the Poynting flux
dominated accretion disk.  One of the purposes of this work is to
see how it depends on the magnetic field.

In the  force-free limit\cite{BZ}, \begin{equation} F_{\mu\nu}
J^\nu =0
        \label{ff},\end{equation}
we get
        \begin{equation} \vec{E}=-{\bar{\omega}\over{\alpha}} \Omega_F ~
        e_{\hat{\phi}} \times \vec{B}^p \label{fff2} ,\end{equation}
in the Schwarzschild background.  $\Omega_F$ is the angular
velocity of the magnetic surface which rotates rigidly in an
axisymetric and stationary state \cite{TPM}. Then the   accretion
equations driven by the Poynting flux  can be rewritten   as
 \ba
(\partial_r u_0) \m +r\w \Omega_F B^{\hat{\phi}}
B^{\hat{\theta}}=0
        \label{tc4} ,\\
        (\partial_r u_\phi) \m -r\w
        B^{\hat{\phi}}B^{\hat{\theta}}=0 \label{tc5} ,\\
   {\m\over 2\pi r^2} g^{rr} {u^0\over u^r}(\partial_r u_\phi)
        \left[ \Omega_D+{\partial_r u_0 \over
        \partial_r u_\phi} \right] \nonumber \\
         -{\sqrt{\Delta}\over 2\pi
        r^2}B^{\hat{r}}B^{\hat{\theta}}\left({\w^2\over\alpha^2}
        \Omega_F^2-1 \right) =0  \label{tc6} .\ea
From eq.(\ref{tc4}) and (\ref{tc5}), one obtains an interesting
relation \begin{equation} {\partial_r u_0\over\partial_r
u_\phi}=-\Omega_F \label{omegaf}.\end{equation} which has no
explicit dependence on the field configuration. It seems to imply
that $\Omega_F$ is determined essentially by the dynamics of the
disk. However one should note that the dynamics itself is governed
not only by the gravity but also by the electromagnetic field as
well.

\subsection{Stream equation}

The configuration of the ordered magnetic field around the
accretion disk has been discussed both in analytical and numerical
studies. The developments  of the ordered magnetic field in the
disk and the Poynting outflow from the disk have been studied
numerically by many authors\cite{ga,lop,fendt}. For the strong
enough electromagnetic field around the compact object, the
force-free magnetosphere can be established. In the
non-relativistic formulation, Blandford\cite{blandford} suggested
an axisymmetric and stationary electromagnetic field configuration
around  an accretion disk on the equatorial plane.  The poloidal
field configuration for a black hole is known to satisfy a second
order elliptical differential equation called Grad-Shafranov
equation\cite{beskin} or a stream equation\cite{MT} for the stream
function $\Psi$ and current $I$. The poloidal and toroidal
components of the  magnetic field can be written in terms of
$\Psi$ and $I$ respectively:
 \ba \vec{B}^P &=& {1\over2\pi\w}\nabla \Psi \times e_{\hat{\phi}} , \nn
 \\
\quad B^{\hat{\phi}} &=& -{2I\over \w \alpha} \label{bpt}.   \ea
 Possible types of solutions in a  force-free
magnetosphere has been discussed recently in the relativistic
formulation\cite{ghosh,llk} .

In a flat background, the stream equation becomes simpler one
given by
        \ba \partial^2_r\Psi +{\sin\theta\over r^2}
        \partial_\theta\left({1\over\sin\theta}\partial_\theta \Psi \right)
        -\Omega_F \sin^2\theta\partial_r \left(r^2 \Omega_F \partial_r
        \Psi \right) \nn
        \\-\Omega_F\sin\theta\partial_\theta\left(\sin\theta\Omega_F\partial_\theta\Psi\right)
        =-16\pi^2
        I {dI\over d\Psi} \label{sf}.\ea
Blandford\cite{blandford} suggested two types of solution of
eq.(\ref{sf}) depending on the types of surface, where  magnetic
field lines rotating with angular velocity $\Omega_F$ lie on. In a
upper  hemisphere, the magnetic field of paraboloidal type cutting
the disk on the equatorial plane is given by
    \ba B^{\hat{r}} &=& { C \over 2r
        \{1+\Omega^2_F r^2 (1-\cos\theta)^2 \}^{1/2}} ,\nn\\
        B^{\hat{\theta}}&=&{-C(1- \cos\theta)\over 2 r \sin\theta
        \{1+\Omega^2_F r^2 (1-\cos\theta)^2 \}^{1/2}} ,\nn\\
        B^{\hat{\phi}}&=&{- C \Omega_F(1-\cos\theta)\over \sin\theta
        \{1+\Omega^2_F r^2 (1-\cos\theta)^2\}^{1/2}},\label{bpara}\ea
and the solution of  hyperboloidal type is given by
\begin{widetext}
\ba B^{\hat{r}}&=&{B_0 R^2 (1-\cos\theta)^2\over \{\sin^2\theta
R^2 -r^2 (1-\cos\theta)^4\}\{R^2-\Omega_F^2 r^4
(1-\cos\theta)^4\}^{1/2}} \nn \\ &\times& \left[(1-\cos\theta)
\{\sin^2\theta-(1-\cos\theta)^2\}^{1/2}
+\{R^2-r^2(1-\cos\theta)^2\}^{1/2}\cos\theta\right] \nn \\
B^{\hat{\theta}}&=& {B_0 R^2 (1-\cos\theta)^2\over \{\sin^2\theta
R^2 -r^2 (1-\cos\theta)^4\}\{R^2-\Omega_F^2 r^4
(1-\cos\theta)^4\}^{1/2}} \nn \\ &\times& \left[(1-\cos\theta)
\{\sin^2\theta-(1-\cos\theta)^2\}^{1/2} \cos\theta/
\sin\theta-\{R^2-r^2(1-\cos\theta)^2\}^{1/2}\sin\theta\right]
\nn\\ B^{\hat{\phi}}&=& {-B_0 R r (1-\cos\theta)^2\Omega_F
\{\sin^2\theta R^2-r^2(1-\cos\theta)^4\}\over \sin\theta
\{\sin\theta^2 R^2 -r^2 (1-\cos\theta)^4\} \{R^2-\Omega_F^2 r^4
(1-\cos\theta)^4\}^{1/2}} \label{bhyper}, \ea \end{widetext}
 where
$C$ and $B_0$ represent the scale of field strength  and $R$
corresponds to the shape parameter for the hyperboloidal
configuration. It is interesting to note that $\Omega_F$ is an
arbitrary function of $r$ and $\theta$.

For the accretion flow,  the magnetic field  on the equatorial
plane, $\theta = \pi/2$, is relevant.  Among the accretion
equations the implications of eq.(\ref{tc6}) are quite different
for those two configurations. For the hyperboloidal type,
$B^{\hat{r}}=0$ on the equatorial plane in eq.(\ref{bhyper}) and
eq.(\ref{tc6}) reads
 \ba  \Omega_D - \Omega_F   =0  \label{tc6h}, \ea
which may be considered to be  the perfect-conductor boundary
condition on the accretion disk.  For the paraboloidal
configuration, eq.(\ref{tc6}) can be written as
  \begin{equation} \m {u^0\over u^r}(\partial_r u_\phi)
\left[\Omega_D-\Omega_F \right]=-{C^2\over 4r(1+\Omega_F^2
)}\left(r^2 \Omega_F^2-1\right). \end{equation} Hence one cannot
expect $\Omega_D=\Omega_F$ unless $v_F(=r\Omega_F)=1$ everywhere,
which seems to be unphysical.  It is  natural to suppose that the
velocity of the magnetic field line $v_F(=r\Omega_F)$ does not
exceed the speed of light.  Then the r.h.s of the equation is
positive. Since the term $\m (u^0/u^r)\partial_r u_\phi$ of the
equation is negative, the term $[\Omega_D-\Omega_F]$ has to be
negative such that
        \begin{equation} \Omega _F>\Omega_D .\end{equation}
With a black hole at the center, however,  the flat background is
no more valid and the field configuration is expected to  be
changed accordingly. Since there is no known analytic solution in
the background of  a black hole, we have to rely on the  numerical
calculation. The detailed numerical calculation will be discussed
in the next section particularly for the paraboloidal
configuration in the Schwarzschild background.

In summary,  we have  a simple picture of  accretion flow  in an
axisymmetric, stationary and force-free magnetosphere, which is
described by the above four equations  in the frame work of a
simplified two-dimensional disk model on the equatorial plane. The
physical variables for this simple system\footnote{$u^{\theta}=0$
and  only two of  $u^0, u^r, u^{\phi}$ are independent because of
the velocity normalization, $u^{\mu}u_{\mu} = -1$.} are  $u^0,
u^r, u^{\phi}$, $\Psi$, $I$  and $\dot{M}_+$. Among those
variables only  $\dot{M}_+$ is independent of the disk location
and we take it as an input parameter in  this work.  Then we can
solve the equations for  this simple disk model to see the effect
of the magnetic field on the accretion flow as well as to find the
field configuration subject to the steady state accretion flow
driven by the magnetic field itself.

\section{Numerical results for a paraboloid-type configuration in
Schwarzschild background}

In a flat background, we can find  $I$ and $\Psi$ corresponding to
the paraboloidal type configuration given by
        \begin{equation} I(X)={\pm C\Omega_F X\over
        2(1+\Omega^2_F X^2)^{1/2}},\label{i}\end{equation}
        \begin{equation} {d\Psi\over d X}={\pi C \over(1+\Omega^2_F X^2)^{1/2}} \label{psi},\end{equation}
where \ba X = r(1\pm \cos \theta). \ea  The  explicit forms for
the magnetic fields  are just those  given in the previous
section, eq.(\ref{bpara}). When $\Omega_F$ is constant, we can
easily integrate $d\Psi/dX$ of eq.(\ref{psi}) to get explicit form
of $\Psi$ as
        \begin{equation} \Psi(X) = {\pi C\over\Omega_F} \sinh^{-1} (\Omega_F X). \end{equation}

In the Schwarzschild background, the stream equation is given by
\begin{widetext}
        \ba \partial_r \left\{\left(1-{2M\over r}\right)\partial_r \Psi
        \right\}+{\sin\theta\over r^2}
        \partial_\theta\left({1\over\sin\theta}\partial_\theta \Psi \right)
        -\Omega_F \sin^2\theta\partial_r \left(r^2 \Omega_F \partial_r
        \Psi \right) \nn \\
        -{\Omega_F\over 1-{2M\over
        r}}\sin\theta\partial_\theta\left(\sin\theta\Omega_F\partial_\theta\Psi\right)
        =-{16\pi^2
        I {dI\over d\Psi}\over \left(1-{2M\over r}\right)}\label{st2}.\ea
\end{widetext} Compared to the case with flat background we do not
have analytical forms like eq.(\ref{i}) and (\ref{psi}).  When
$\Omega_F$ and $I$ are vanishing, eq.(\ref{st2}) becomes
        \begin{equation}\partial_r \left\{\left(1-{2M\over r}\right)\partial_r \Psi
        \right\}+{\sin\theta\over r^2}
        \partial_\theta\left({1\over\sin\theta}\partial_\theta \Psi
        \right)=0 .\end{equation}
One of the solutions, $\Psi_0$, suggested by Blandford and Znajek
\cite{BZ} is given by
        \begin{equation} \Psi_0 =\pi C X
        \label{psi0},\end{equation}
where \begin{equation} X\equiv r(1\mp\cos\theta)+2M
        (1\pm\cos\theta)\{1-\log(1\pm\cos\theta)\}, \end{equation}
and upper/lower sign refers to the upper/lower hemisphere.  One
can easily see that it reduces to the paraboloidal field lines for
$M \rightarrow 0$ or $ r \gg M$. The upper (lower) sign
corresponds to the northern (southern) hemisphere. It is
interesting to note that the discontinuity of $B^{\hat{r}}$ on the
disk surface$(\theta=\pi/2)$ defines the toroidal surface current
$K^{\hat{\phi}}$
        \begin{equation} K^{\hat{\phi}}={C\over 4\pi r}
        \label{current-boundary},\end{equation}
which is the same as in the flat background\cite{blandford}.

For non-vanishing $\Omega_F$ and $I$, we consider a solution for
which  the shape of the magnetic surface are the same as the
magnetic surface defined by $\Psi_0$\cite{m84}. Thus $\Psi , I$
and $\Omega_F $ are assumed to  depend  on X. We suppose that the
derivative of $\Psi$ has the same form as in the flat background,
eq.(\ref{psi}). That is, we take an Ansatz such that
        \begin{equation} {d\Psi\over d X}={\pi C \over(1+\Omega^2_F
        X^2)^{1/2}}.
        \label{ansatz}\end{equation}
On the equatorial plane, $X=r+2M$, we can construct magnetic
fields on the disk surface  as follows:
        \ba B^{\hat{r}} &=& {\pm C \over 2r (1+\Omega^2_F X^2 )^{1/2}} \nn, \\
        B^{\hat{\theta}}&=&{-C\over 2 r (1+\Omega^2_F X^2 )^{1/2}}
        \left(1-{2M\over r}\right)^{1/2} \nn, \\
        B^{\hat{\phi}}&=&  {-2I\over r}\left(1-{2M\over r}\right)^{-1/2}
        \label{ms3}. \ea
Using  eq.(\ref{ansatz}), eq.(\ref{st2})  can be written as
     \begin{equation} {4\pi C \Omega_F\over(1+\Omega^2_F X^2)^{3/2}} { \Delta\over(X-2M)^2}
    =-16\pi^2 I {dI\over
    d\Psi} \label{i2},\end{equation}
where
\begin{widetext}
    \ba \Delta={d\Omega_F\over dX}(X-3M)
    (4M^3-6M^2X+5MX^2-X^3) + \Omega_F
    \left\{10M^3-18M^2X+8MX^2  -X^3+3M^2X(X-2M)^2\Omega^2_F\right\}.\ea
\end{widetext}
It can be considered to be an equation for $I$ on the disk
surface$(\theta=\pi/2)$ provided $\Omega_F$ is given.  In contrast
to the flat background case, we do not have the analytic form for
$I$ analogous to eq.(\ref{i}).

The accretion equations can then be written as follows,
          \begin{equation} (\partial_r u_0) \m +{C^2 \Omega_F\over(1+\Omega_F^2 X^2)}\tilde{I} =0
          \label{tc7} ,\end{equation}
          \begin{equation} (\partial_r u_\phi)\m -{C^2\over(1+\Omega_F^2
    X^2)}\tilde{I} =0 \label{tc8}, \end{equation}
    \begin{equation} \m {u^0\over u^r}(\partial_r
    u_\phi) \left[ {u^\phi\over u^0}-\Omega_F \right]
   +{C^2\over 4r(1+\Omega_F^2 X^2)}\left({r^2 \Omega_F^2\over 1-{2M\over
    r}}-1\right) =0 \label{tc9}, \end{equation}
where  a dimensionless quantity $\tilde{I}$ for the upper
hemisphere is defined by
 \begin{equation} \tilde{I} \equiv (1+\Omega_F^2 X^2)^{1/2} {I\over C} ~~ .\end{equation}

Four basic equations, eq.(\ref{i2}) - eq.(\ref{tc9}), governing
the accretion flow under the influence of paraboloidal-type
configuration  are solved numerically\cite{jp}. We get the radial
variations of $u_0$, $u_{\phi}$, $I$ and $\Omega_F$ on the disk
with dimensionless variable $\mt= \dot{M}_+ /C^2 $ as an input
parameter in this work. It should be noted that the dimensionless
accretion rate parameter $\mt$ is the relative magnitude of the
accretion rate to $C^2$ which determines the strength-squared of
the magnetic field.

For numerical calculations, we need  a set of  initial values to
start with. We take an initial point $r_0$ far from the horizon
such that we can make use of the Newtonian approach in the initial
step of the numerical calculation. We take $r_0$ to be much
greater than inner most stable radius\cite{st}
        \begin{equation} r_0 \gg 6M.\end{equation}
In this asymptotic region we can take initial values as those for
the flat background approximation. At first, we  choose initial
values for $u^0$ and $u^\phi$ to be
        \begin{equation} u^0 =u^0_K ,\quad u^\phi = \eta \,u^\phi_K,\end{equation}
where  $u^0_K$ and   $u^\phi_K$ are those for the Keplerian orbit.
We  take the initial value of $u^\phi$ to be smaller than that of
the Keplerian orbit to prevent $u^r$ from becoming an imaginary
number, $\eta<1$. Plugging the initial values of $u^0$ and
$u^\phi$ into  eq.(\ref{tc8}) and eq.(\ref{tc9}), the initial
value of $\Omega_F$ can be determined. And the initial value of
$\tilde{I}$ is then determined by making use of
 of flat-background relation in
eq.(\ref{i}):
        \begin{equation} \tilde{I} = {1\over2} \Omega_F r
        \label{iflat}.\end{equation}
After fixing these initial values, it is straightforward to
execute the numerical calculation.

For numerical calculation, we  consider  two cases of the
accretion-rate parameters, [a] $\mt=0.7$ and [b] $\mt=3$ and take
initial point to be $r_0=50M$. To get some physical idea on the
values of $\mt$, let us take $C$ to be 2.78$\times 10^{19}$
gauss$\cdot$cm for a solar mass black hole as an  example. Then
the magnetic field strength becomes $\sim 10^{12}$ gauss at
$r=r_0$ which might be be compatible with the astrophysical
phenomena accompanied with strong magnetic field, like magnetar
and  GRB \cite{lwb}. Then the accretion rates are given by
$\m=1.8\times 10^{28}$g/s \,and $7.7\times 10^{28}$g/s \,for
$\mt=0.7$ and $3$ respectively.

\begin{figure*}

\includegraphics{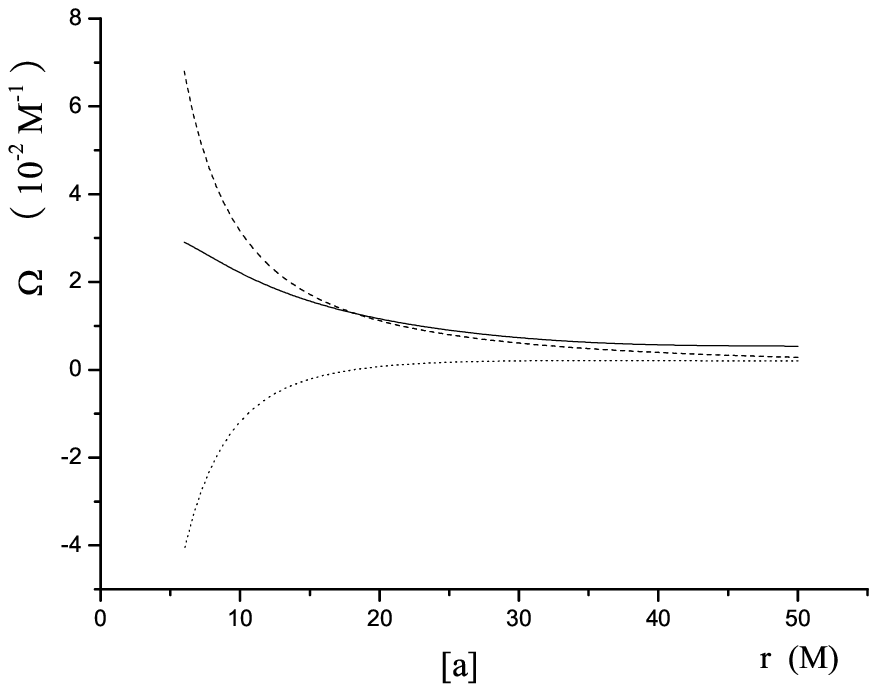} \includegraphics{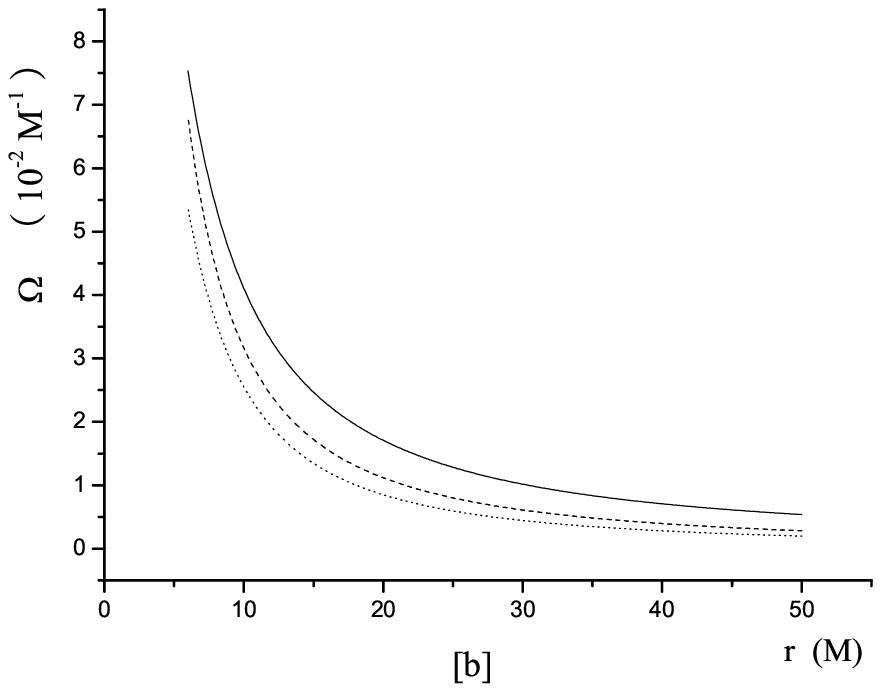}

\caption{\label{fig:omega1}The angular velocities  of the
   magnetic surface $\Omega_F$, accretion disk $\Omega_D$ and  Keplerian orbit $\Omega_K$ are shown
   in   solid,  dotted  and  dashed lines respectively for two cases of accretion rate:
     [a] $\mt = 0.7$ and [b] $\mt = 3$.}
\end{figure*}

We continue the numerical calculation inwardly up to $r=6M$ which
is the radius of the innermost stable circular orbit in the
Schwarzschild black hole at the center. As shown in Figure
\ref{fig:omega1}, the angular velocities  of the magnetic field
lines($\Omega_F$) are found to be different from either  the
Keplerian angular velocity($\Omega_K$) or the disk angular
velocity($\Omega_D)$. It implies that the two-dimensional disk
model may not be a good approximation for the accretion disk of
perfect conductor. As expected in the previous section for the
velocity of the magnetic field line $v_F(=r\Omega_F)$ less than
the  speed of light,   $\Omega_F$ is found to be larger than
$\Omega_D$.

For  a smaller  accretion rate,  $\mt = 0.7$, $\Omega_D$ changes
its sign at $r \sim 20M$ as shown in Figure \ref{fig:omega1}[a].
It corresponds to changing its sense of rotation.  It is because
for a given strength of the magnetic field, the slope of angular
momentum ($\partial_r u_\phi$) is proportional to the inverse of
$\dot{M}_+$ as can be seen in eq.(\ref{tc5}). Hence the angular
momentum of the accretion flow decreases more rapidly for smaller
$\m$ and   there is a chance of reverting the direction of
rotation. But for larger $\mt$, the slope of change is not high
enough to change the sense of rotation,  Figure
\ref{fig:omega1}[b].

\begin{figure*}
\includegraphics{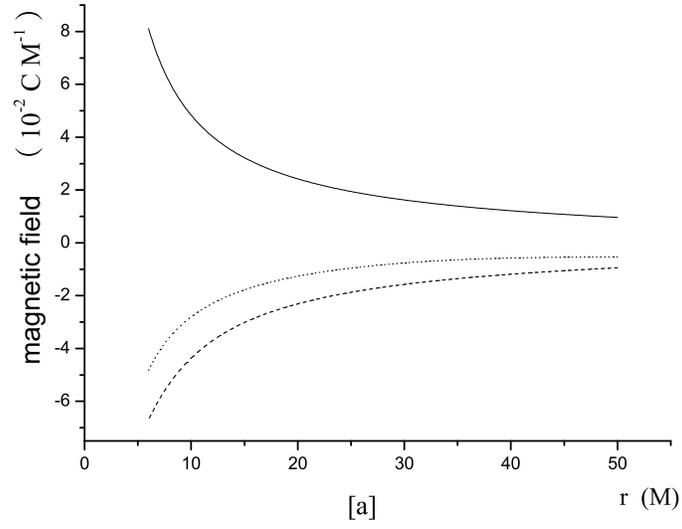} \includegraphics{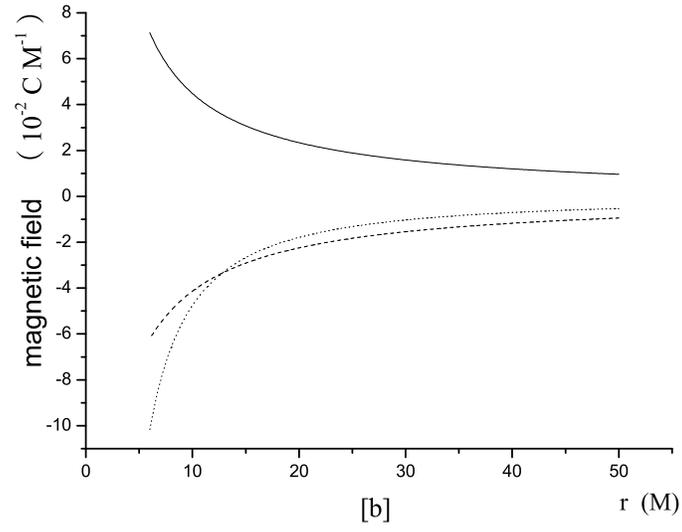}

\caption{\label{fig:bfield1} The strength of the magnetic field in
the upper hemisphere: the axial component $B^{\hat{\phi}}$,
$B^{\hat{r}}$ and $B^{\hat{\theta}}$ are represented by the
dotted, solid and dashed lines respectively.}
\end{figure*}

\begin{figure*}
\includegraphics{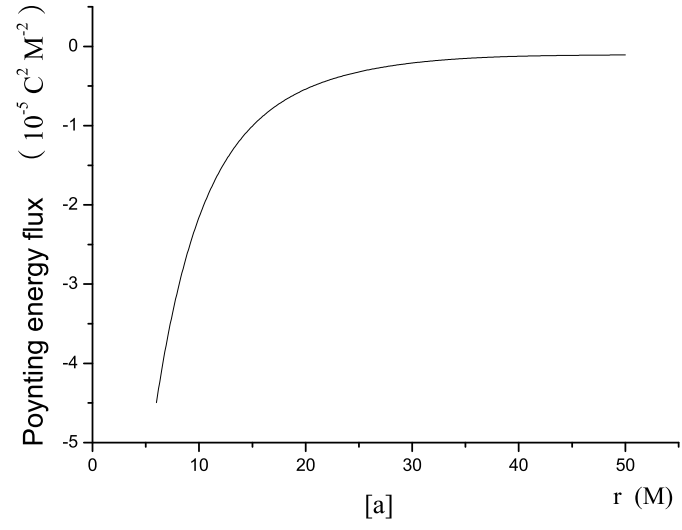} \includegraphics{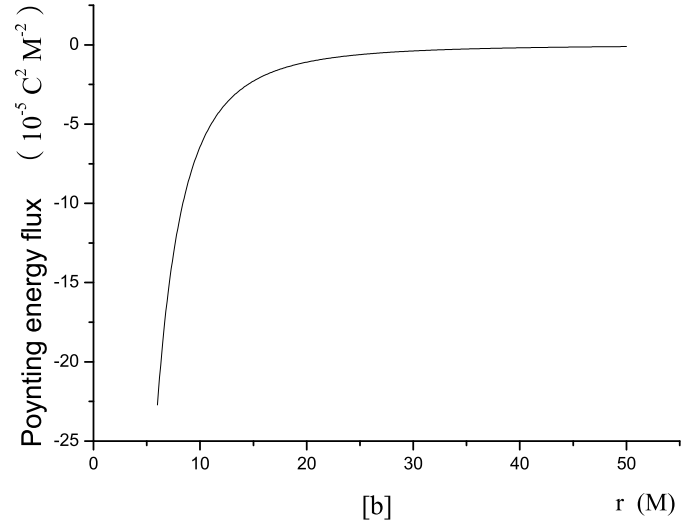}

\caption{\label{fig:energyflux1} Poynting energy flux, ${\mathcal
E}^{\hat{\theta}}$, in the upper hemisphere.}
\end{figure*}

We obtain also the numerical solutions for $\Psi$ and $I$ of the
stream equation to get the electromagnetic field on the disk. It
is found that the strength of the magnetic field is increasing as
$r$ goes near the inner edge  as shown in Figure
\ref{fig:bfield1}. One can see that  the axial component
$B^{\hat{\phi}}$ is sensitive on $\mt$ while $B^{\hat{r}}$ and
$B^{\hat{\theta}}$ are not sensitive to $\mt$. The surface
currents $K^i$ are defined as discontinuities of the magnetic
fields across the disk and  they show similar behaviors as those
of the magnetic field. Since the magnetic field  as well as the
surface current are increasing as $r$ gets smaller, it is
naturally expected that the Poynting flux increases substantially
as $r$ approaches to the center as shown in Figure
\ref{fig:energyflux1}. Although the numerical calculation does not
go beyond $r<6M$, it may indicate a possible electromagnetic jet
structure  near the inner edge of the disk.

\section{Discussion}

In this work, we investigate the accretion flow dominated by
Poynting flux using  a simplified model of a two-dimensional disk.
The two-dimensional accretion disk in a force-free magnetosphere
is described by the basic four equations, three accretion
equations and one stream equation. We observe that the condition
of the stationary accretion flow (constant $\dot{M}_+$) gives rise
to nontrivial  constraints on the form of $\Omega_F$ and on the
configuration of the ambient electromagnetic field. It is found
that it depends strongly on the types of field. In the flat
background limit,  we find that the hyperbloidal type
configuration suggested by Blandford  can accommodate the
solutions with $\Omega_F = \Omega_D$, which is consistent with an
accretion disk of perfect conductor.  However  for the
paraboloidal  type configuration,  $\Omega_F$ is expected to be
greater than  $\Omega_D$. Hence one can guess that a
two-dimensional flow with paraboloidal type configuration might be
relevant for an accretion disk not in a perfect-conducting state.

To get more details on the dynamics of the accretion flow, we try
to find numerical solutions particularly for the paraboloidal type
configuration in Schwarzschild background. The overall dynamics of
the disk is found to depend essentially on the ratio of the
accretion rate to the field strength-squared, $\mt$. When it is
sufficiently small, the inner part of the accretion disk is found
to be changing its sense of rotation to match the angular momentum
balance with respect to the Poynting flux.

One of the interesting results of this work is to demonstrate
clearly how $\Omega_F$ is different from either $\Omega_D$ or
$\Omega_K$ under the influence of the strong magnetic field. Since
the numerical calculation in this work  is extended only up to the
inner stable radius for the Schwarzschild black hole, the
relativistic effect is found not to be dominant such that the
discussion in a flat background limit does not change
substantially.  Hence the observation of $\Omega_F = \Omega_D$ for
hyperboloidal type configuration in a two-dimensional accretion
flow is expected to be valid even with Schwarzschild black hole at
the center.

We demonstrate in this work  that how the accretion disk provides
relevant boundary conditions, which  is  important to find the
solutions of the stream equation.  In this work it is possible
basically because we have a mathematically well-defined disk
surfaces in this simplified model as in the case of horizon.
However   it would be interesting to see whether  any realistic
accretion disk model can provide the relevant boundary conditions
for the stream equation, to which our analysis can be applied.

In order to simulate  the strong relativistic effect it is
necessary to extend the present study of the two-dimensional
accretion flow to that in the Kerr background. However as the
inner edge of the accretion disk gets very near/or beyond the
ergo-sphere for a rapidly rotating black hole, the magnetic
coupling between the black hole and the accretion
disk\cite{gammie,krolik,wang} becomes important and the simple
picture in this work should be modified substantially.

\begin{acknowledgments}
This work was supported by the research fund of Hanyang
University(HY-2003-1). \end{acknowledgments}


\newpage

\end{document}